\documentclass[onecolumn,aps]{revtex4}

\usepackage{graphicx}
\newcommand{\be}{\begin{equation}}
\newcommand{\ee}{\end{equation}}
\newcommand{\ben}{\begin{eqnarray}}
\newcommand{\een}{\end{eqnarray}}

\begin{document}

\title{Critical behaviour of the compactified $\lambda \phi^4$ theory}

\author{L.M. Abreu, C. de  Calan, A. P. C. Malbouisson\footnote{Permanent
address:
CBPF/MCT, Rua Dr. Xavier Sigaud, 150, Rio de Janeiro RJ, Brazil}}
\address{{\it Centre de Physique Th{\'e}orique, Ecole Polytechnique, 91128
Palaiseau, France}}
\author{J. M. C. Malbouisson\footnote{Permanent address: Instituto de
F{\'\i}sica, Universidade Federal da Bahia, 40210-340, Salvador, BA, Brazil}}
\address{{Department of Physics, University of Alberta, Edmonton,
AB, T6G 2J1, Canada}}
\author{A. E. Santana}
\address{{Instituto de F{\'\i}sica, Universidade de Bras{\'\i}lia, 70910-900,
Bras{\'\i}lia-DF, Brasil}}

\begin{abstract}

$\;$\\
\\
\noindent {\bf Abstract}\\

We investigate the critical behaviour of the $N$-component Euclidean $\lambda \phi^4$ model
at leading order in $\frac{1}{N}$-expansion. We consider it in three situations: confined between two parallel
planes a distance $L$ apart from one another, confined to an infinitely long
cylinder having a square cross-section of area $A$ and to a cubic box
of volume $V$. Taking the mass term in the form $m_{0}^2=\alpha(T - T_{0})$,
we retrieve Ginzburg-Landau models which are supposed to describe
samples of a material undergoing a phase transition, respectively in the form
of a film, a wire and of a grain, whose bulk transition temperature
($T_{0}$) is known. We obtain equations for the critical
temperature as  functions of $L$ (film), $A$ (wire),
$V$ (grain) and of $T_{0}$, and determine the limiting sizes sustaining the
transition.\\

\noindent PACS number(s): 11.10.Jj; 11.10.Kk; 11.15.Pg\\
\\

\end{abstract}
\maketitle

\noindent {\bf 1. Introduction}\\
\\

Models with fields confined in spatial dimensions play important roles
both in field theory and in quantum mechanics. Relevant examples are
the Casimir effect and superconducting films, where confinement is
carried on by appropriate boundary conditions. For Euclidean field
theories, imaginary time and the spatial coordinates are treated
exactly on the same footing, so that an extended Matsubara formalism
can be applied for dealing with the breaking of invariance along any
one of the spatial directions.\\

Relying on this fact, in the present work we discuss the critical
behaviour of the Euclidean $\lambda\varphi^{4}$ model compactified
in one, two and three spatial dimensions. We implement the
spontaneous symmetry breaking by taking the bare mass coefficient
in the Lagrangean parametrized as $m_{0}^2=\alpha(T - T_{0})$,
with $\alpha > 0$ and the parameter $T$ varying in an interval
containg $T_0$. With this choice, considering the system confined
between two paralell planes a distance $L$ apart from one another,
in an infinitely long square cylinder with cross-section area
$A=L^2$, and in a cube of volume $V=L^3$, in dimension $D=3$, we
obtain Ginzburg-Landau models describing phase transitions in
samples of a material in the form of a film, a wire and a grain,
respectively, $T_0$ standing for the bulk transition temperature.
Such discriptions apply to physical circumstances where no gauge
fluctuations need to be considered. \\

We start presenting a recapitulation of the general procedure developed in Ref. 
\cite{Ademir} to treat the massive
$(\lambda \varphi ^{4})_ {D}$ theory in Euclidean space,
compactified in a $d$-dimensional subspace, with $d\leq D$. This
permits to extend to an arbitrary subspace some results in the
literature for finite temperature field theory \cite{NAdolfo1} and
for the behaviour of field theories in presence of spatial
boundaries \cite{JMario,Ademir,FAdolfo}. We shall consider the
vector $N$-component $(\lambda \varphi ^{4})_{D}$ Euclidean theory
at leading order in $\frac{1}{N}$, thus allowing for
non-perturbative results, the system being submitted to the
constraint of compactification of a $d$-dimensional subspace.
After that, besides the review of the situation $d=1$ (already studied in Refs. \cite{MMS1}), 
we extend the investigation to the two other particularly interesting cases of $d=2$ and $d=3$. These three situations above mentioned correspond respectively to the system confined between paralell
planes (a film), confined to an infinitely long cylinder of square
cross-section (a wire)
and to a finite cubic box (a grain).\\

For these situations, in the framework of the Ginzburg-Landau
model we derive equations for the critical temperature as a
function of the confining dimensions. For a film, we show that the
critical temperature decreases linearly with the inverse of the
film thickness while, for a square cross-section wire and for a
cubic grain, we obtain that the critical temperature decreases
linearly with the inverse of the square root of the cross-section
area $A$ and with the inverse of the cubic root of the grain
volume $V$, respectively. In all cases, we are able to calculate
the minimal system size (thickness, cross-section area, or volume)
below which the phase transition does not take place.\\
\\

\noindent {\bf 2. The compactfied model}\\
\\

In this Section we review the analytical methods of compactification of the $N$-component Euclidean $\lambda \phi^4$ model developed in Ref. \cite{Ademir}. 
We consider the model described by the Hamiltonian density,
\begin{equation}
{\cal {H}}=\frac{1}{2}\partial _{\mu }\varphi _{a}\partial ^{\mu }\varphi
_{a}+\frac{1}{2}\bar{m}_{0}^{2}\varphi _{a}\varphi _{a}+
\frac{\lambda }{N}(\varphi_{a}\varphi _{a})^{2},  \label{Lagrangeana}
\end{equation}
in Euclidean $D$-dimensional space, confined to a $d$-dimensional
spatial rectangular box of sides $L_{j}$, $j=1,2,...,d$.  In the
above equation $\lambda $ is the {\it renormalized}  coupling
constant, $\bar{m}_{0}^{2}$ is a boundary-modified mass parameter
depending on $\{L_i\}\;i=1,2...d$, in such a way that
\begin{equation}
\lim_{\{L_i\}\rightarrow\infty}\bar{m}_{0}^{2}(L_1,...,L_d) = m_{0}^{2}(T) 
\equiv \alpha \left( T - T_0 \right),
\label{m0}
\end{equation}
$m_{0}^{2}(T)$ being the constant mass parameter present in the usual
free-space Ginzburg-Landau model. In Eq.~(\ref{m0}), $T_0$ represents the
bulk transition temperature. Summation over repeated ``color''
indices $a$ is assumed. To simplify the notation in the following
we drop out the color indices, summation over them being
understood in field products. We will work in the approximation of
neglecting boundary corrections to the coupling constant. A
precise definition of the boundary-modified mass parameter will be
given later for the situation of $D=3$ with $d=1$, $d=2$ and $d=3$,
corresponding respectively to a film of thickness $L_{1}$, to a wire of 
rectangular section $L_{1}\times L_{2}$ and to a grain of volume 
$L_{1}\times L_{2}\times L_{3}$.\\

We use Cartesian coordinates ${\bf r}=(x_{1},...,x_{d},{\bf z})$,
where ${\bf z}$ is a $(D-d)$-dimensional vector, with
corresponding momentum ${\bf k}=(k_{1},...,k_{d}, {\bf q})$, ${\bf
q}$ being a $(D-d)$-dimensional vector in momentum space. Then the
generating functional of correlation functions has the form,
\begin{equation}
{\cal Z}=\int {\cal D}\varphi ^{\dagger }{\cal D}\varphi \exp \left(
-\int_{0}^{{\bf L}}d^{d}r\int d^{D-d}{\bf z}\;{\cal H}(\varphi,\nabla
\varphi)
\right) \,,  \label{part}
\end{equation}
where ${\bf L}=(L_{1},...,L_{d})$, and we are allowed to introduce
a generalized Matsubara prescription, performing the following
multiple replacements (compactification of a $d$-dimensional
subspace),
\begin{equation}
\int \frac{dk_{i}}{2\pi }\rightarrow \frac{1}{L_{i}}\sum_{n_{i}=-\infty
}^{+\infty }\;;\;\;\;\;\;\;k_{i}\rightarrow \frac{2n_{i}\pi }{L_{i}}
\;,\;\;i=1,2,...,d.
\label{Matsubara1}
\end{equation}
A simpler situation is the system confined simultaneously between two
parallel planes a distance $L_1$ apart from one another normal to the
$x_1$-axis and two other parallel planes, normal to the $x_2$-axis
separated by a distance $L_2$ (a ``wire" of rectangular section).\\

We start from the well known expression for the one-loop contribution
to the zero-temperature effective potential\cite{IZ},
\begin{equation}
U_{1}(\varphi _{0})=\sum_{s=1}^{\infty }\frac{(-1)^{s+1}}{2s}\left[ 12
\lambda \varphi _{0}^{2}\right] ^{s}\int
\frac{d^{D}k}{(2\pi)^D}\frac{1}{(k^{2}+m^{2})^{s}}.\label{potefet0}
\end{equation}
where $m$ is the physical mass and $\varphi _{0}$ is the normalized
vacuum expectation value of the field (the classical field).  In the
following, to deal with dimensionless quantities in the regularization
procedures, we introduce parameters
\begin{equation}
c=\frac{m}{2\pi\mu},\;\;b_{i}=\frac{1}{L_{i}\mu},\;\;
g=\frac{\lambda}{4\pi^2\mu^{4-D}},\;\;\phi_{0}^2 =
\frac{\varphi_{0}^{2}}{\mu^{D-2}},
\label{semdim}
\end{equation}
where  $\mu $ is a mass scale. In terms of these parameters and
performing the replacements (\ref{Matsubara1}), the one-loop
contribution to the effective potential can be written in the
form,
\begin{equation}
U_{1}(\phi _{0},b_{1},...,b_{d})=\mu^{D}\, b_{1}\cdots b_{d}
\;\sum_{s=1}^{\infty }\frac{(-1)^{s}}{2s} \left[ 12g\phi
_{0}^{2} \right]^{s} \sum_{n_{1},...,n_{d}=-\infty }^{+\infty }\int
\frac{d^{D-d}q^{\prime}}{(b_{1}^{2}n_{1}^{2}+\cdots +b_{d}^{2}n_{d}^{2}+c^{2}+
{\bf q}^{\prime 2})^{s}} \;,
\label{potefet1}
\end{equation}
where ${\bf q}^{\prime}={\bf q}/2\pi\mu$ is dimensionless. Using a
well-known dimensional regularization formula \cite{Zinn} to
perform the integration over the ($D-d$) non-compactfied momentum
variables, we obtain
\begin{equation}
U_{1}(\phi _{0},b_{1},...,b_{d})=\mu ^{D}\, b_{1}\cdots
b_{d}\;\sum_{s=1}^{\infty }f(D,d,s)\left[ 12 g\phi
_{0}^{2} \right]^{s} A_{d}^{c^{2}}
\left(s-\frac{D-d}{2};b_{1},...,b_{d}\right),
\label{potefet2}
\end{equation}
where
\begin{equation}
f(D,d,s)=\pi ^{(D-d)/2}\frac{(-1)^{s+1}}{2s\Gamma (s)}\Gamma
(s-\frac{D-d}{2})
\end{equation}
and
\begin{eqnarray}
A_{d}^{c^{2}}(\nu ;b_{1},...,b_{d})& = & \sum_{n_{1},...,n_{d} = -\infty}
^{+\infty }(b_{1}^{2}n_{1}^{2}+\cdots +b_{d}^{2}n_{d}^{2}+c^{2})^{-\nu }
   =  \frac{1}{c^{2\nu }}+2\sum_{i=1}^{d}\sum_{n_{i}=1}^{\infty}
(b_{i}^{2}n_{i}^{2}+c^{2})^{-\nu}  \nonumber \\
 & & +\;  2^{2}\sum_{i<j=1}^{d}\sum_{n_{i},n_{j}=1}^{\infty
}(b_{i}^{2}n_{i}^{2}+b_{j}^{2}n_{j}^{2}+c^{2})^{-\nu }+\cdots  +
2^{d}\sum_{n_{1},...,n_{d}=1}^{\infty }(b_{1}^{2}n_{1}^{2}+\cdots
+b_{d}^{2}n_{d}^{2}+c^{2})^{-\nu }.  \label{zeta}
\end{eqnarray}\\

Next we can proceed generalizing to several dimensions the mode-sum
regularization prescription described in Ref. \cite{Elizalde}. This
generalization has been done in \cite{Ademir} and we briefly describe
here its principal steps.  From the identity,
\begin{equation}
\frac{1}{\Delta ^{\nu }}=\frac{1}{\Gamma (\nu )}\int_{0}^{\infty
}dt\;t^{\nu of this work
-1}e^{-\Delta t},
\end{equation}
and using the following representation for Bessel functions of
the third kind, $K_{\nu }$,
\begin{equation}
2(a/b)^{\frac{\nu }{2}}K_{\nu }(2\sqrt{ab})=\int_{0}^{\infty }dx\;x^{\nu
-1}e^{-(a/x)-bx},  \label{K}
\end{equation}
we obtain after some rather long but straightforward manipulations
\cite{Ademir},
\begin{eqnarray}
A_{d}^{c^{2}}(\nu ;b_{1},...,b_{d}) &=&\frac{2^{\nu -\frac{d}{2}+1}\pi
^{2\nu -\frac{d}{2}}}{b_{1}\cdots b_{d}\,\Gamma (\nu )}\left[ 2^{\nu
- \frac{d}{2}-1}\Gamma \left(\nu -\frac{d}{2}\right) \left( 2 \pi c
\right)^{d-2\nu } + 2\sum_{i=1}^{d}\sum_{n_{i}=1}^{\infty }\left(\frac{n_{i}}
{2\pi c b_{i}}\right)^{\nu - \frac{d}{2}}K_{\nu
-\frac{d}{2}}\left( \frac{2\pi c n_{i}}{b_i} \right)\right.  \nonumber \\
&&\left.+\cdots +2^{d}\sum_{n_{1},...,n_{d}=1}^{\infty }
\left(\frac{1}{2\pi c}
\sqrt{\frac{n_{1}^{2}}{b_{1}^{2}} + \cdots + \frac{n_{d}^{2}}{b_{d}^{2}}}
\right)^{\nu - \frac{d}{2}}
K_{\nu - \frac{d}{2}} \left( 2\pi c
\sqrt{\frac{n_{1}^{2}}{b_{1}^{2}} + \cdots + \frac{n_{d}^{2}}{b_{d}^{2}}}
\right)\right] .
\label{zeta4}
\end{eqnarray}
Taking $\nu =s-(D-d)/2$ in Eq.~(\ref{zeta4}) and inserting it in
Eq.~(\ref{potefet2}), we obtain the one-loop correction to the effective
potential in $D$ dimensions with a compactified $d$-dimensional
subspace in the form (recovering the dimensionful parameters)
\begin{eqnarray}
U_{1}(\varphi _{0},L_{1},...,L_{d}) &=&\sum_{s=1}^{\infty }\left[ 12
g\phi
_{0}^{2} \right]^{s} h(D,s)\left[ 2^{s-\frac{D}{2}-2}\Gamma
(s-\frac{D}{2}) m^{D-2s}
+\sum_{i=1}^{d}\sum_{n_{i}=1}^{\infty }\left(\frac{m}{L_{i}n_{i}}
\right)^{\frac{D}{2}-s} K_{\frac{D}{2}-s}\left( mL_{i}n_{i}\right)\right.
\nonumber \\
&&+\left.
2\sum_{i<j=1}^{d}\sum_{n_{i},n_{j}=1}^{\infty}\left(\frac{m}{
\sqrt{L_{i}^{2}n_{i}^{2}+L_{j}^{2}n_{j}^{2}}}\right)^{\frac{D}{2}-s}
K_{\frac{D}{2}-s}\left( m\sqrt{L_{i}^{2}n_{i}^{2}+L_{j}^{2}n_{j}^{2}}\right)
+\cdots \right.
\nonumber
\\
&&+\left. 2^{d-1}\sum_{n_{1},...,n_{d} = 1}^{\infty
}\left(\frac{m}{\sqrt{L_{1}^{2}n_{1}^{2}+\cdots
+L_{d}^{2}n_{d}^{2}}}\right)^{\frac{D}{2}-s}
K_{\frac{D}{2}-s}\left( m\sqrt{L_{1}^{2}n_{1}^{2}+\cdots
+L_{d}^{2}n_{d}^{2}}\right)\right] ,
 \label{potefet3}\end{eqnarray}
with
\begin{equation}
h(D,s)=\frac{1}{2^{D/2+s-1}\pi^{D/2}}\frac{(-1)^{s+1}}{s\Gamma (s)}\; .
\label{h}
\end{equation}

Criticality is attained when the inverse squared correlation
length , $\xi ^{-2}(L_{1},...,L_{d},\varphi _{0})$, vanishes in the
large-$N$ gap equation,
\begin{eqnarray}
\xi ^{-2}(L_{1},...,L_{d},{\bf \varphi }_{0})
&=&\bar{m}_{0}^{2}+12\lambda
\varphi _{0}^{2} + \frac{24\lambda }{L_{1}\cdots L_{d}}
\nonumber  \\
&& \times\sum_{n_{1},...,n_{d}=-\infty
}^{\infty }\int
\frac{d^{D-d}q}{(2\pi )^{D-d}}\;\frac{1}{{\bf q}^{2}+(\frac{2\pi n_{1}}{
L_{1} })^{2}+...+(\frac{2\pi n_{d}}{L_{d}})^{2}+\xi ^{-2}(L_{1},...,L_{d},
\varphi _{0})},
\label{gap}
\end{eqnarray}
where ${\bf \varphi }_{0}$ is the normalized vacuum expectation
value of the field (different from zero in the ordered phase). In
the disordered phase, $\varphi _{0}$ vanishes and the inverse
correlation length equals the physical mass, given below by
Eq.~(\ref{massa}).
Recalling  the condition,
\begin{equation}
\left. \frac{\partial ^{2}}{\partial \varphi_{0} ^{2}}U(D,L_1 ,L_2)\right|
_{\varphi _{0}=0}=m^{2}\,  \label{renorm1}
\end{equation}
where $U$ is the sum of the tree-level and one-loop contributions
to the effective potential (remembering that at the large-$N$
limit it is enough to take the one-loop contribution to the mass),
we obtain
\begin{eqnarray}
m^{2}(L_{1},...,L_{d}) &=&\bar{m}_{0}^{2}(L_{1},...,L_{d})
+\frac{24\lambda }{(2\pi)^{D/2}}\left[
\sum_{i=1}^{d}\sum_{n_{i}=1}^{\infty }\left(\frac{m}{L_{i}n_{i}}
\right)^{\frac{D}{2}-1}K_{\frac{D}{2}-1}\left( mL_{i}n_{i}\right)\right.
\nonumber \\
&&+\left. 2\sum_{i<j=1}^{d}\sum_{n_{i},n_{j}=1}^{\infty }\left(\frac{m}{
\sqrt{L_{i}^{2}n_{i}^{2}+L_{j}^{2}n_{j}^{2}}}\right)^{\frac{D}{2}-1}
K_{\frac{D}{2}-1}\left( m\sqrt{L_{i}^{2}n_{i}^{2}+L_{j}^{2}n_{j}^{2}}\right)
+\cdots \right.
\nonumber
\\
&&+\left. 2^{d-1}\sum_{n_{1},...,n_{d} = 1}^{\infty }\left(\frac{m}{
\sqrt{L_{1}^{2}n_{1}^{2}+\cdots +L_{d}^{2}n_{d}^{2}}}\right)^{\frac{D}{2}-1}
K_{\frac{D}{2}-1}\left( m\sqrt{L_{1}^{2}n_{1}^{2}+\cdots
+L_{d}^{2}n_{d}^{2}}\right)\right] \, .  \label{massa}
\end{eqnarray}
Notice that, in writing Eq.~(\ref{massa}), we have suppressed the parcel
$2^{-\frac{D}{2}-1}\Gamma (1-\frac{D}{2}) m^{D-2}$ from its square bracket,
the parcel that emerges from the first term in the square bracket of
Eq.~(\ref{potefet3}). This expression, which does not depend explicitly on
$L_{i}$, diverges for $D$ even due to the poles of the gamma function;
in this case, this parcel is subtracted to get
a renormalized mass equation. For $D$ odd, $\Gamma\left( 1-\frac{D}{2}\right)$
is finite but we also subtract this term (corresponding to a finite
renormalization) for sake of uniformity; besides, for $D \geq 3$, the factor
$m^{D-2}$ does not contribute in the criticality.\\

The vanishing of Eq.~(\ref{massa}) defines criticality for our
compactified system. We claim that, taking $d=1$, $d=2$, and $d=3$ with $D=3$,
we are able to describe respectively the critical behaviour of
samples of materials in the form of films, wires and grains.
Notice that the parameter $m$ in the right hand side of
Eq.(\ref{massa}) is the boundary-modified mass
$m(L_{1},...,L_{d})$, which means that Eq.(\ref{massa}) is a
self-consistency equation, a very complicated modified
Schwinger-Dyson equation for the mass, not soluble by algebraic
means. Nevertheless, as we will see in the next sections, a
solution is possible at criticality, which allows us to obtain a
closed formula for the boundary-dependent critical temperature.\\
\\

\noindent {\bf 3. Critical behaviour for films}\\
\\

We now consider the simplest particular case of the compactification of
only one spatial dimension, the system confined between two parallel
planes a distance $L$ apart from one another. This case has been already considered in Ref. \cite{MMS1}, and we also analyze it here for completeness. Thus, from Eq.~(\ref{massa}), 
taking $d=1$, we get in the disordered
phase
\begin{equation}
m^{2}(L) =\bar{m}_{0}^{2}(L)+\frac{24\lambda}{(2\pi
)^{D/2}}
\sum_{n=1}^{\infty }\left( \frac{m}{nL }\right)^{\frac{D}{2}-1}
K_{\frac{D}{2}-1}(n L m)  \, ,
\label{massafilme}
\end{equation}
where $L$ ($=L_1$) is the separation between the planes, the
film thickness. If we limit ourselves to the neighbourhood of criticality
($m^{2}\approx 0$) and consider $L$ finite and sufficiently small, we may
use an asymptotic formula for small values of the argument of
Bessel functions,
\begin{equation}
K_{\nu }(z)\approx \frac{1}{2}\Gamma (|\nu| )\left( \frac{z}{2}\right)
^{-|\nu|
}\;\;\;\;\;\;(z\approx 0) \; ,  \label{KK}
\end{equation}
and Eq.~(\ref{massafilme}) reduces, for $D>3$, to
\begin{equation}
m^{2}(L)\approx  \bar{m}_{0}^{2}(L)+\frac{6\lambda}
{\pi^{D/2} L^{D-2} }\Gamma \left(\frac{D}{2}-1\right)\zeta(D-2)
\label{filmecr1}
\end{equation}
where $\zeta (D-2)$ is the Riemann $zeta$-function, defined for
${\rm Re}\{ D-2\} > 1$ by the series
\begin{equation}
\zeta (D-2)=\sum_{n=1}^{\infty }\frac{1}{n^{D-2}}.
\end{equation}
It is worth mentioning that for $D=4$,
taking $m^{2}(L)=0$ and making the appropriate changes
($L\rightarrow \beta$, $\lambda \rightarrow \lambda/4 !$),
Eq.(\ref{filmecr1}) is {\it formally identical} to the
high-temperature (low values of $\beta$) critical equation obtained 
in Ref.~\cite{DJ}, thus providing a check of our calculations.\\

For $D=3$, Eq.(\ref{Tcr1}) can be made  physically  meaningful  by
a regularization procedure as follows. We consider the analytic
continuation of the $zeta$-function, leading to a meromorphic function
having only one simple pole at $z=1$, which satisfies the reflection formula
\begin{equation}
\zeta (z)=\frac{1}{\Gamma (z/2)}\Gamma (\frac{1-z}{2})\pi ^{z-\frac{1}{2}
}\zeta (1-z)\;.  \label{extensao}
\end{equation}
Next, remembering the formula,
\begin{equation}
\lim_{z\rightarrow 1}\left[ \zeta (z)-\frac{1}{z-1}\right] =\gamma\; ,
\label{extensao1}
\end{equation}
where $\gamma\approx 0.5772$ is the Euler-Mascheroni constant, we
define the L-dependent bare mass for $D\approx 3$, in such a way that
the pole at $D=3$  in Eq.(\ref{filmecr1}) is suppressed, that is we take
\begin{equation}
\bar{m}_{0}^{2}(L) \approx M - \frac{1}{(D-3)}\frac{6\lambda}{\pi L}\, ,
\label{massrenfilme}
\end{equation}
where $M$ is independent of $D$. To fix the finite term, we make the simplest 
choice satisfying (\ref{m0})
\begin{equation}
M = m_{0}^{2}(T) = \alpha \left( T - T_{0}\right) \, ,
\end{equation} 
$T_0$ being the bulk critical temperature. 
In this case, using Eq.~(\ref{massrenfilme}) in Eq.~(\ref{filmecr1}) and 
taking the limit as $D\rightarrow 3$, the $L$-dependent renormalized mass term 
in the vicinity of criticality becomes
\begin{equation}
m^{2}(L) \approx \alpha \left( T - T_{c}(L) \right) \, ,
\label{MRF}
\end{equation}
where the modified, $L$-dependent, transition temperature is given by
\begin{equation}
T_{c}(L) = T_0 - C_1 \frac{\lambda}{\alpha L}\; ,
\label{Tcr1}
\end{equation}
$L$ being the thickness of the film, with the constant $C_1$ given by
\begin{equation}
C_1 = \frac{6\gamma}{\pi} \approx 1.1024 \; . \label{C1}
\end{equation} 
From this equation, we see that for $L$ smaller than
\begin{equation}
L_{\rm min} = C_1 \frac{\lambda}{\alpha T_0} \; ,
\end{equation}
$T_c(L)$ becomes negative, meaning that the transition does not
occurs \cite{MMS1}.\\
\\

\noindent {\bf 4. Critical behaviour for wires}\\
\\

We now focus on the situation where two spatial dimensions are compactified.
From Eq.~(\ref{massa}), taking $d=2$, we get (in the disordered phase)
\begin{eqnarray}
m^{2}(L_1 ,L_2) &=&\bar{m}_{0}^{2}(L_1 ,L_2) 
+ \frac{24\lambda}{(2\pi)^{D/2}}
\left[\sum_{n=1}^{\infty }(\frac{m}{nL_1 })^{\frac{D}{2}-1}K_{\frac{D}{2}
-1}(nL_{1} m)+\sum_{n=1}^{\infty
}(\frac{m}{nL_2})^{\frac{D}{2}-1}K_{\frac{D}{
2 }-1}(nL_{2}m)\right.  \nonumber  \label{mDyson} \\
&&+\left. 2\sum_{n_{1},n_{2}=1}^{\infty }(\frac{m}{\sqrt{L_{1}
^{2}n_{1}^{2}+L_{2}^{2}n_{2}^{2}}})^{\frac{D}{2}-1}K_{\frac{D}{2}-1}(m\sqrt{
L_{1} ^{2}n_{1}^{2}+L_{2}^{2}n_{2}^{2}})\right].
\end{eqnarray}
If we limit ourselves to the neighborhood of criticality, $
m^{2}\approx 0$, and taking both $L_1$ and $L_2$ finite and sufficiently
small, we may use Eq.(\ref{KK}) to rewrite Eq.~(\ref{mDyson}) as
\begin{eqnarray}
m^{2}(L_1 ,L_2) & \approx & \bar{m}_{0}^{2}(L_1 ,L_2) 
+ \frac{6\lambda}
{\pi^{D/2}}\Gamma \left(\frac{D}{2}-1\right) \nonumber \\
&&\times \left[ \left(\frac{1}{L_{1}^{D-2}}+\frac{1}{L_{2}^{D-2}}\right)\zeta
(D-2)+2E_{2}\left(\frac{D-2}{2};L_{1},L_{2}\right) \right] \;,
\label{mDysoncr}
\end{eqnarray}
where  $E_{2}\left(\frac{D-2}{2};L_{1},L_{2}\right)$ is the generalized
(multidimensional) Epstein $zeta$-function defined by
\begin{equation}
E_{2}\left(\frac{D-2}{2};L_{1},L_{2}\right) =
\sum_{n_{1},\,n_{2}=1}^{\infty} \left[ L_{1}
^{2}n_{1}^{2}+L_{2}^{2}n_{2}^{2}\right]^{-\left(
\frac{D-2}{2}\right)}\, , \label{Z}
\end{equation}
for ${\rm Re}\{ D \} > 3$.\\

As mentioned before, the Riemann $zeta$-function $\zeta (D-2)$ has
an analytical extension to the whole complex $D$-plane, having an
unique simple pole (of residue $1$) at $ D=3$. One can also
construct analytical continuations (and recurrence relations) for
the multidimensional Epstein functions which permit to write them
in terms Kelvin and Riemann $zeta$ functions. To start one
considers the analytical continuation of the Epstein-Hurwitz
$zeta$-function given by \cite{Elizalde}
\begin{equation}
\sum_{n=1}^{\infty}\left( n^2 + p^2 \right)^{-\nu} = -\frac{1}{2} p^{-2\nu}
+ \frac{\sqrt{\pi}}{2 p^{2\nu -1}\Gamma (\nu)} \left[
\Gamma\left( \nu-\frac{1}{2} \right)
+ 4\sum_{n=1}^{\infty} (\pi p n)^{\nu-\frac{1}{2}}
K_{\nu-\frac{1}{2}}(2\pi p n) \right] \; . \label{EHfunc}
\end{equation}
Using this relation to perform one of the sums in (\ref{Z}) leads
immediately to the question of which sum is firstly evaluated. As
it is done in Ref. \cite{Kirsten}, whatever the sum one chooses to
perform firstly, the manifest $L_{1} \leftrightarrow L_{2}$
symmetry of Eq.~(\ref{Z}) is lost; in order to preserve this
symmetry, we adopt here a symmetrized summation. Generalizing the
prescription introduced in \cite{Ademir}, we consider the
multidimensional Epstein function defined as the symmetrized
summation
\begin{equation}
E_{d}\left(\nu;L_{1},...,L_{d}\right) = \frac{1}{d !}
\sum_{\sigma} \sum_{n_1 = 1}^{\infty} \cdots \sum_{n_d =
1}^{\infty} \left[ \sigma_{1}^{2} n_{1}^{2} + \cdots +
\sigma_{d}^{2} n_{d}^{2} \right]^{\, - \nu} \; , \label{EfuncS}
\end{equation}
where $\sigma_{i}=\sigma(L_{i})$, with $\sigma$ running in the set
of all permutations of the parameters $L_1,...,L_d$, and the
summations over $n_1,...,n_d$ being taken in the given order.
Applying (\ref{EHfunc}) to perform the sum over $n_d$, one gets
\begin{eqnarray}
E_{d}\left(\nu;L_{1},...,L_{d} \right) & = & -
\,\frac{1}{2\, d} \sum_{i=1}^{d} E_{d-1}\left( \nu;...,{\widehat
{L_{i}}},...\right) \nonumber \\
 & & +\, \frac{\sqrt{\pi}}{2\, d\, \Gamma(\nu)} \Gamma\left( \nu - \frac{1}{2}
 \right) \sum_{i=1}^{d} \frac{1}{L_i} E_{d-1}
 \left( \nu-\frac{1}{2};...,{\widehat {L_{i}}},... \right)
 + \frac{2\sqrt{\pi}}{d\, \Gamma(\nu)} W_d \left( \nu-\frac{1}{2},L_1,...,L_d
 \right)\; , \label{Wd}
\end{eqnarray}
where the hat over the parameter $L_{i}$ in the functions
$E_{d-1}$ means that it is excluded from the set $\{L_{1},...,L_{d}\}$ 
(the others being the $d-1$ parameters of $E_{d-1}$), and
\begin{equation}
W_d\left(\eta;L_1,...,L_d\right) = \sum_{i=1}^{d} \frac{1}{L_i}
\sum_{n_1,...,n_d=1}^{\infty} \left( \frac{\pi n_i}{L_i \sqrt{(
\cdots + {\widehat {L_i^2 n_i^2}} + \cdots )}} \right)^{\eta}
K_{\eta}\left( \frac{2\pi n_i}{L_i} \sqrt{( \cdots + {\widehat
{L_i^2 n_i^2}} + \cdots )} \right) \; , \label{WWd}
\end{equation}
with $(\cdots + {\widehat {L_i^2 n_i^2}} + \cdots)$ representing
the sum $\sum_{j=1}^{d}L_{j}^{2}n_{j}^{2} \, - \,
L_{i}^{2}n_{i}^{2}$. In particular, noticing that 
$E_{1}\left(\nu; L_j\right)=L_{j}^{-2\nu}\zeta(2\nu)$, one finds
\begin{eqnarray}
E_{2}\left(\frac{D-2}{2};L_{1} ^{2},L_{2}^{2}\right) & = &
-\frac{1}{4}\left(\frac{1}{L_{1}^{D-2}} + \frac{1}{L_{2}^{D-2}}\right)
\zeta (D-2)  \nonumber \\
&&+\;\frac{\sqrt{\pi }\Gamma (\frac{D-3}{2})}{4\Gamma (\frac{D-2}{2})}\left(
\frac{1}{L_{1} L_{2}^{D-3}}+\frac{1}{L_{1} ^{D-3}L_{2}}\right) \zeta (D-3)
+\frac{\sqrt{\pi }}{\Gamma (\frac{D-2}{2})}
W_{2}\left(\frac{D-3}{2};L_{1},L_{2}\right)\;,
\label{Z1}
\end{eqnarray}

Using the above expression, the Eq.~(\ref{mDysoncr}) can be rewritten as
\begin{eqnarray}
m^{2}(L_1,L_2) & \approx & 
\bar{m}_{0}^{2}(L_1,L_2) + \frac{3\lambda}{\pi ^{D/2}}\left[
 \left(\frac{1}{L_{1}^{D-2}} + \frac{1}{L_{2}^{D-2}}\right)
\Gamma \left(\frac{D-2}{2} \right)
\zeta (D-2) \right. \nonumber \\
 & & + \left.  \sqrt{\pi} \left( \frac{1}{L_{1}L_{2}^{D-3}} +
  \frac{1}{L_{1}^{D-3}L_{2}} \right) \Gamma \left(
  \frac{D-3}{2} \right) \zeta(D-3) +\, 2 \sqrt{\pi}  
W_{2}\left(\frac{D-3}{2};L_{1},L_{2}\right)
\right] \, .  \label{EQC1}
\end{eqnarray}
This equation presents no problems for $3<D<4$ but, for 
$D=3$, the first and second terms between brackets of Eq.
(\ref{EQC1}) are divergent due to the $\zeta$-function and
$\Gamma$-function, respectively.  We can deal with divergences
remembering  the property in Eq. (\ref{extensao1}) and using the
expansion of $\Gamma(\frac{D-3}{2})$  around $D=3$,
\begin{equation}
\Gamma(\frac{D-3}{2})\approx \frac{2}{D-3}+\Gamma'(1)
\label{GD3}
\end{equation}
$\Gamma '(z)$ standing for the derivative of the $\Gamma$-function
with respect to $z$. For $z=1$ it coincides with the Euler
digamma-function $\psi(1)$, which has the particular value
$\psi(1)=-\gamma$. We notice however, that differenlty from the
case treated in the previous section, where it was necessary a
renormalization procedure, here the two divergent terms generated
by the use of formulas (\ref{extensao1}) and (\ref{GD3}) cancel exactly
between them. No renormalization is needed. Thus, for $D=3$, taking the bare 
mass given by $\bar{m}_{0}^{2}(L_1,L_2) = \alpha \left( T - T_0 \right)$, we 
obtain the renormalized boundary-dependent mass term in the form
\begin{equation}
m^{2}(L_1,L_2) \approx \alpha \left( T - T_{c}(L_1,L_2) \right) \, , 
\label{MRW} 
\end{equation}
with the boundary-dependent critical temperature given by
\begin{equation}
T_{c}(L_1,L_2) = T_0 - \frac{9 \lambda \gamma}{2\pi
\alpha}\left(\frac{1}{L_{1}}+\frac{1}{L_2}\right)  -\frac{6\lambda }{\pi
\alpha} W_{2}(0;L_1,L_2) \, ,
\label{Tcr}
\end{equation}
where
\begin{equation}
W_{2}(0;L_1,L_2) = \sum_{n_1,n_2 = 1}^{\infty} \left\{
\frac{1}{L_1} K_{0}\left( 2\pi\frac{L_{2}}{L_{1}}n_{1}n_{2}\right) + 
\frac{1}{L_2} K_{0}\left( 2\pi\frac{L_{1}}{L_{2}}n_{1}n_{2}\right) 
\right\}.
\label{W2L12}
\end{equation}\\

The quantity $W_{2}(0;L_1,L_2)$, appearing in Eq.(\ref{Tcr}),
involves complicated double sums, very
difficult to handle for $L_1 \neq L_2$; in particular, it is not
possible to take limits such as $L_i \rightarrow \infty$. For this
reason we will restrict ourselves to the case $L_1=L_2$. 
For a  wire with square cross-section, we have $L_1=L_2=L=\sqrt{A}$ and
Eq.~(\ref{Tcr}) reduces to
\begin{equation}
T_{c}(A) = T_0 - C_{2} \frac{\lambda}{\alpha \sqrt{A}} \; ,
\label{Tcr2}
\end{equation}
where $C_2$ is a constant given by
\begin{equation}
C_2 = \frac{9\gamma}{\pi} + \frac{12}{\pi}
\sum_{n_1,n_2=1}^ {\infty}K_{0}(2\pi n_{1}n_{2}) \approx 1.6571 \; .
\label{w}
\end{equation}\\

We see that the critical temperature of the square wire depends on
the bulk critical temperature and the Ginzburg-Landau parameters
$\alpha$ and $\lambda$ (which are characteristics of the material
constituting the wire), and also on the area of its
cross-section. Since $T_c$ decreases linearly with the inverse of
the side of the square, this suggests that there is a minimal area
for which $T_c(A_{\rm min})=0$,
\begin{equation}
A_{\rm min} = \left( C_{2} \frac{\lambda}{\alpha T_{0}}\right)^{2} \; ;
\label{Amin}
\end{equation}
for square wires of cross-section areas smaller than this value,
in the context of our model the transition should be suppressed.
On topological grounds, we expect that (apart from appropriate
coefficients) our result should be independent of the
cross-section shape of the wire, at least for cross-sectional
regular polygons.\\
\\

\noindent {\bf 5. Critical behaviour for grains}\\
\\

We now turn our attention to the case where all three spatial dimensions are 
compactified, corresponding to the system confined in a box of sides 
$L_{1},L_{2},L_{3}$. Taking $d=3$ in Eq.~(\ref{massa}) and using 
Eq.~(\ref{KK}), we obtain (for sufficiently small $L_{1},L_{2},L_{3}$ and in 
the neighbourhood of classicality, $m^2\approx 0$)
\begin{eqnarray}
m^{2}(L_1,L_2,L_3) & \approx & \bar{m}_{0}^{2}(L_1,L_2,L_3) \nonumber \\
 & & +\, \frac{6\lambda }
{\pi^{D/2}}\Gamma \left(\frac{D-2}{2}\right)
\left[ \sum_{i=1}^{3} \frac{\zeta(D-2)}{L_{i}^{D-2}}  
+2 \sum_{i<j=1}^{3}
  E_{2}\left( \frac{D-2}{2};L_{i},L_{j} \right)
+ 4E_{3} \left( \frac{D-2}{2};L_{1},L_{2},L_{3}
  \right) \right] \; , \nonumber \\
 & & \label{ECG1}
\end{eqnarray}
where 
$E_{3}( \nu;L_{1},L_{2},L_{3}) =
\sum_{n_1,n_2,n_3=1}^{\infty }
\left[ L_{1}^{2}n_{1}^{2}+L_{2}^{2}n_{2}^{2}+L_{3}^{2}n_{3}^{2}\right]
^{-\,\nu}$
and the functions $E_2$ are given by Eq.~(\ref{Z1}).\\

The analytical structure of the function
$E_{3}\left(\frac{D-2}{2};L_{1},L_{2},L_{3}\right)$
can be obtained from the general symmetrized recurrence
relation given by Eqs.~(\ref{Wd},\ref{WWd}); explicitly, one has
\begin{eqnarray}
E_{3}\left(\frac{D-2}{2};L_{1},L_{2},L_{3}\right) & = &
-\frac{1}{6}\sum_{i<j=1}^{3}
  E_{2}\left(\frac{D-2}{2};L_{i},L_{j}\right) 
+ \frac{\sqrt{\pi}\Gamma (\frac{D-3}{2})}{6\Gamma (\frac{D-2}{2})}
\sum_{i,j,k=1}^{3}\frac{(1+\varepsilon_{ijk})}{2}
\frac{1}{L_{i}} E_2 \left(\frac{D-2}{2};L_{j},L_{k}\right)
\nonumber \\
& & +\frac{2\sqrt{\pi}}{3\Gamma(\frac{D-2}{2})} \, 
W_{3}\left(\frac{D-3}{2};L_{1},L_{2},L_{3}\right),
\label{E3}
\end{eqnarray}
where $\varepsilon_{ijk}$ is the totally antisymmetric symbol and the 
function $W_3$ is a particular case of Eq.~(\ref{WWd}). Using 
Eqs.~(\ref{Z1}) and (\ref{E3}), the boundary dependent mass can be written as 
\begin{eqnarray}
m^{2}(L_1,L_2,L_3) & \approx & \bar{m}_{0}^{2}(L_1,L_2,L_3)+\frac{6\lambda}
{\pi^{D/2}} \left[ \frac{1}{3} \Gamma \left(\frac{D-2}{2}\right)
   \sum_{i=1}^{3}\frac{1}{L_{i}^{D-2}}\zeta(D-2) \right. \nonumber \\
 & & +\, \frac{\sqrt{\pi}}{6} \zeta(D-3)\sum_{i<j=1}^{3}
  \left( \frac{1}{L_{i}^{D-3}L_{j}}
+ \frac{1}{L_{j}^{D-3}L_{i}} \right) \Gamma\left( \frac{D-3}{2} \right) 
\, +\, \frac{4\sqrt{\pi}}{3}\sum_{i<j=1}^{3}
W_{2}\left(\frac{D-3}{2};L_{i},L_{j}\right) \nonumber \\   
& &   +\, \frac{\pi}{6} \zeta(D-4)
\Gamma \left(\frac{D-4}{2}\right)
\sum_{i,j,k=1}^{3} \frac{(1+\varepsilon_{ijk})}{2}
\frac{1}{L_{i}}\left( \frac{1}{L_{j}^{D-4}L_{k}}
    +\frac{1}{L_{k}^{D-4}L_{j}} \right) \nonumber \\
& & \left.  +\, \frac{2\pi}{3}
\sum_{i,j,k=1}^{3} \frac{(1+\varepsilon_{ijk})}{2}
\frac{1}{L_{i}}W_{2}\left(\frac{D-4}{2};L_{j},L_{k}\right)
\, +\, \frac{8\sqrt{\pi}}{3}W_{3}\left(\frac{D-3}{2};L_{1},L_{2},L_{3}\right) 
\right]
\label{ECG2}
\end{eqnarray}
The first two terms in the square bracket of Eq.~(\ref{ECG2}) diverge as 
$D\rightarrow 3$ due to the poles of the $\Gamma$ and $\zeta$-functions. 
However, as it happens in the case of wires, using Eqs.~(\ref{extensao1})
and (\ref{GD3}) it can be shown that these divergences cancel exactly 
one another. After some simplifications, for $D=3$, the boundary dependent mass
(\ref{ECG2}) becomes
\begin{eqnarray}
m^{2}(L_1,L_2,L_3) & \approx & \bar{m}_{0}^{2}(L_1,L_2,L_3)+\frac{6\lambda}
{\pi} \left[ \frac{\gamma}{2}
   \sum_{i=1}^{3}\frac{1}{L_{i}} +
\frac{4}{3}
\sum_{i<j=1}^{3} W_{2}(0;L_{i},L_{j})
+ \frac{\pi}{18}
\sum_{i,j,k=1}^{3} \frac{(1+\varepsilon_{ijk})}{2}  
\frac{L_{i}}{L_{j}L_{k}}  \right. \nonumber \\
& & \left. +\, \frac{2\sqrt{\pi}}{3}
\sum_{i,j,k=1}^{3} \frac{(1+\varepsilon_{ijk})}{2} 
\frac{1}{L_{i}}W_{2}\left(-\frac{1}{2};L_{j},L_{k}\right)
+\frac{8}{3}W_{3}(0 ;L_{1}, L_{2}, L_{3}) \right] \; .
\label{ECG3}
\end{eqnarray}\\

As before, since no divergences need to be suppressed, we can take the bare 
mass given by $\bar{m}_{0}^{2}(L_1,L_2,L_3) = \alpha (T - T_0)$ and rewrite 
the renormalized mass as $m^{2}(L_1,L_2,L_3) \approx \alpha 
\left( T - T_{c}(L_1,L_2,L_3) \right)$. The expression of $T_{c}(L_1,L_2,L_3)$ 
can be easily obtained from Eq.~(\ref{ECG3}), but it is a very complicated 
formula, involving multiple sums, which makes almost impossible a general 
analytical study for arbitrary parameters $L_1,L_2,L_3$; thus, we restrict 
ourselves to the situation where $L_1=L_2=L_3=L$, corresponding to a cubic 
box of volume $V = L^3$. In this case, the boundary dependent critical 
temperature reduces to
\begin{equation}
T_{c}(V) =  T_{0} - C_3 \frac{\lambda}{\alpha V^{1/3}} \; ,
\label{ECG4}
\end{equation}
where the constant $C_3$ is given by (using that $K_{- \frac{1}{2}}(z) = 
\sqrt{\frac{\pi}{2 z}}e^{-z}$)
\begin{eqnarray}
C_3 & = & 1 + \frac{9\gamma}{\pi} + 
\frac{12}{\pi} \sum_{n_1,n_2 =1}^{\infty}\frac{e^{-2 \pi n_{1}n_{2}}}{n_{1}} 
\nonumber \\
 & & +\, \frac{48}{\pi} \sum_{n_1,n_2 =1}^{\infty} K_{0}(2\pi n_{1}n_{2}) + 
\frac{48}{\pi} \sum_{n_1,n_2,n_3=1}^{\infty} K_{0}\left( 2\pi n_{1}
\sqrt{n_{2}^{2}+n_{3}^{2}} \right) \approx 2.7657 \; . \label{C3}
\end{eqnarray}
One sees that the minimal volume of the cubic grain sustaining the transition 
is
\begin{equation}
V_{\rm min} = \left( C_{3} \frac{\lambda}{\alpha T_{0}}\right)^{3} 
\; . \label{Vmin}
\end{equation}\\
\\

\noindent {\bf 6. Conclusions}\\
\\

In this paper we have discussed the spontaneous symmetry breaking of 
the $(\lambda\phi^4)_D$ theory compactified in $d\leq D$ Euclidean 
dimensions, extending some results of Ref.~\cite{Ademir}. 
We have parametrized the bare mass term in the form 
$m_{0}^{2}(T-T_0)$, thus placing the analysis within the Ginzburg-Landau 
framework. We focused in the situations with $D=3$ and $d=1,2,3$, 
corresponding (in the context of condensed matter systems) to films, 
wires and grains, respectively, undergoing phase transitions which may 
be described by (mean-field) Ginzburg-Landau models. This generalizes to more 
compactified dimensions previous investigations on the superconducting 
transition in films, both without \cite{MMS1} and in the presence of a 
magnetic field \cite{AMMS}. In all cases studied here, in the absence of 
gauge fluctuations, we found that the boundary-dependent critical temperature 
decreases linearly with the inverse of the linear dimension $L$,  
$T_c(L)=T_0-C_d\lambda/\alpha L$ where $\alpha$ and $\lambda$ are the 
Ginzgurg-Landau parameters, $T_0$ is the bulk transition temperature and $C_d$ 
is a constant equal to $1.1024$, $1.6571$ and $2.6757$ for $d=1$ (film), 
$d=2$ (square wire) and $d=3$ (cubic grain), respectively. Such behaviour 
suggests the existence of a minimal size of the system below which the 
transition is suppressed.\\

These findings seems to be in {\it qualitative}
agreement with results for the existence of a minimal thickness
for disappearance of superconductivity in films
\cite{Simonin,Quateman,Nb3,Xi}. Also, experimental investigations
in nanowires searching to establish whether there is a limit to
how thin a superconducting wire can be, while retaining its
superconducting character, have also drawn the attention of
researchers; for example, in Ref.~\cite{Tinkham} the behaviour of
nanowires  has been studied. Similar questions have also been
rised concerning the behaviour of superconducting
nanograins \cite{grains,delft}. Nevertheless, an important point to be
emphasized is that our results are obtained in a field-theoretical
framework and do not depend on microscopic details of the material
involved nor account for the influence of manufacturing aspects of the 
sample; in other words, our results emerge solely as a topological effect 
of the compactification of the Ginzburg-Landau model in a subspace. Detailed 
microscopic analysis is required if one attepmts to account quantitatively 
for experimental observations which might desviate from our mean field 
results.\\
\\

\noindent {\bf Acknowledgements}\\
\\

This work was partially supported by CAPES and CNPq (Brazilian agencies).\\


\begin{thebibliography}{99}

\bibitem{Ademir} A.P.C. Malbouisson, J.M.C. Malbouisson, A.E. Santana,
Nucl. Phys. B 631 (2002) 83.

\bibitem{NAdolfo1} C. de Calan, A.P.C. Malbouisson, N.F. Svaiter, Mod.
Phys. lett. A 13 (1998) 1757.

\bibitem{JMario} A.P.C. Malbouisson, J.M.C. Malbouisson, J. Phys. A
Math. Gen. 35 (2002) 2263.

\bibitem{FAdolfo} L. Da Rold, C.D. Fosco, A.P.C. Malbouisson, Nucl.
Phys. B 624 (2002) 485.

\bibitem{MMS1} A.P.C. Malbouisson, J.M.C. Malbouisson, A.E. Santana,
Phys. Lett. A 318 (2003) 406.

\bibitem{IZ} C. Itzykson, J-B. Zuber,
Quantum Field Theory, McGraw-Hill, New York, 1980, p. 451.

\bibitem{Zinn} J. Zinn-Justin, Quantum Field Theory and Critical
Phenomena, Clarendon, Oxford, 1996.

\bibitem{Elizalde} A. Elizalde, E. Romeo, J. Math. Phys. 30 (1989) 1133.

\bibitem{DJ} L. Dolan, R. Jackiw, Phys. Rev. D 9 (1974) 3320,
Eq.~(3.19b).

\bibitem{Kirsten} K. Kirsten, J. Math. Phys. 35 (1994) 459.

\bibitem{AMMS} L.M. Abreu, A.P.C. Malbouisson, J.M.C. Malbouisson,
A.E. Santana, Phys. Rev. B 67 (2003) 212502.

\bibitem{Simonin}  J. Simonin, Phys. Rev. B 33 (1986) 7830.

\bibitem{Quateman}  J. H. Quateman, Phys. Rev. B 34 (1986) 1948.

\bibitem{Nb3}  M. S. M. Minhaj, S. Meepagala, J. T. Chen, L. E.
Wenger, Phys. Rev. B 49 (1994) 15235.

\bibitem{Xi} A.V. Pogrebnyakov, et al, Appl. Phys. Lett. 82 (2003) 4319.

\bibitem{Tinkham} A. Bezryadin, C.N. Lau, and M. Tinkham, Nature 
404 (2000) 971;\\
N. Markovic, C.N. Lau, M. Tinkham, Physica C 387 (2003) 44.

\bibitem{grains} D.C. Ralph, C.T. Black, M. Tinkham,
Phys. Rev. Lett. 76 (1996) 688;\\
D.C. Ralph, C.T. Black, M. Tinkham, Phys. Rev. Lett. 78 (1997) 4087.

\bibitem{delft} J. von Delft, D.C. Ralph, Phys. Rep. 345 (2001) 61;\\
J. von Delft, Ann. Phys. (Leipzig) 10 (2001) 219.

\bibitem{Giordano} N. Giordano, E. Sweetland, Phys. Rev. B 39 (1989) 6455.

\end{thebibliography}
\end{document}